\documentclass[11pt,a4paper]{article}

\usepackage[T1]{fontenc}
\usepackage[margin=1in]{geometry}

\usepackage{amsmath,amssymb,amsthm}

\usepackage{graphicx}
\usepackage{xcolor}

\usepackage[sups]{XCharter}
\usepackage[charter]{mathdesign}
\DeclareMathAccent{\tilde}{\mathalpha}{operators}{"7E}
\usepackage{setspace}

\usepackage{booktabs}

\usepackage{subcaption}

\usepackage{pdfpages}

\usepackage{lineno}

\usepackage[normalem]{ulem}    
\usepackage{soul}

\usepackage[numbers]{natbib}

\usepackage[colorlinks=true,allcolors=black]{hyperref}
\usepackage{cleveref}
\crefname{section}{Section}{Sections}
\crefname{figure}{Figure}{Figures}
\crefname{table}{Table}{Tables}
\crefname{equation}{Equation}{Equations}

\newcommand{\qbar}{\bar{q}}

\newcommand{\wbar}{\bar{w}}

\title{Can We Volunteer Out of the Peer Review Crisis?}
\author{
  Theo Tang\textsuperscript{1},
  Toby Handfield\textsuperscript{2},
  and Julian Garcia\textsuperscript{1,*}
  \\[6pt]
  \small\textsuperscript{1}Department of Data Science and Artificial Intelligence, Monash University, Melbourne, Australia\\
  \small\textsuperscript{2}SOPHIS, Monash University, Melbourne, Australia\\[4pt]
  \small\textsuperscript{*}Corresponding author: \texttt{julian.garcia@monash.edu}
}
\date{\today}

\begin{document}

\frenchspacing

\setstretch{1.3}

\maketitle

\begin{abstract}
The volume of scientific manuscripts is growing faster than the capacity to evaluate them, yet the institutions that govern peer review have remained largely unchanged.
The result is a widening mismatch: reviewer scarcity, noisier assessments, and declining confidence in editorial decisions.
Every scientist wants better reviews, but review quality depends on the total burden, which no single author can shift.
To analyze this tension, we provide a game-theoretic thought experiment: a voluntary lottery in which authors accept a chance of random pre-review rejection, reducing reviewer burden and improving the quality of surviving evaluations.
We show that a Nash equilibrium emerges in which authors voluntarily enter the lottery.
Scientists who care about the literature they read, not just the papers they publish, will opt in, raising the quality of published science for all.
\end{abstract}

\section{Introduction}
\label{sec:introduction}

The peer review system is under mounting strain.
The number of submitted manuscripts has been doubling roughly every decade across most fields~\citep{kovanis2016global}
but the reviewer pool has not kept pace~\citep{hochberg2009tragedy}.
The consequences are predictable: reviewers face ever-larger piles of manuscripts, with less time to devote to each,
and the quality of editorial decisions degrades~\citep{tantithamthavorn2025blended}.
The rise of large language models, which dramatically lower the cost of producing manuscripts,
threatens to accelerate this trend further~\citep{liang2024monitoring}.
Yet the basic architecture of peer review has remained largely unchanged,
creating a growing mismatch between the volume of scientific output and the institutions designed to evaluate it.

Empirical evidence confirms that this degradation is substantial.
In a landmark consistency experiment at NeurIPS (a major machine-learning venue),
two independent review panels disagreed on the fate of roughly one quarter of submissions~\citep{price2014nips,
cortes2021inconsistency}.
A replication at NeurIPS 2021, with five times as many submissions, found the same disagreement rate ---
and showed that making a venue more selective would increase, not decrease,
the arbitrariness of decisions~\citep{beygelzimer2023neurips}.
The pattern is not confined to machine learning: in a study of 4{,}000 ESRC grant proposals,
inter-reviewer correlations were as low as 0.2,
and a single negative review roughly halved a proposal's chances even when other reviews were
strong~\citep{jerrim2020reliable}.
Indeed, peer review has been called a game of chance~\citep{neff2006peerreview}.
These findings raise a basic question: what, if anything,
can be done about the link between the scale of science and the rigorous enforcement of quality?

Recent theoretical work has begun to formalise this problem.
Noisier review encourages weaker papers to be submitted, which increases load and further degrades review quality~\citep{adda2024grantmaking,bergstrom2026screening}.
Game-theoretic and simulation models of peer review~\citep{zollman2023academic, tiokhin2021honest,
thurner2011peerreview,
zhang2022systemlevel} have explored how strategic behaviour by authors
and reviewers interacts with review quality \citep{feliciani2019scoping}, but how much voluntary cooperation among authors could improve the system remains an open question.

At root, this is a type of public goods problem. Scientists want good quality research to be published, but they also want their own research to be published. Individuals maximizing for the latter goal will tend to degrade our collective achievement of the former goal. We can then ask two questions: first, from a social planner's point of view, could we improve the quality of published research by a kind of blanket reduction in participation: do `less' science in order to get `better' science? Second, if scientists are sufficiently motivated by the collective quality of science, could some sort of voluntary scheme to reduce the volume of submissions be sustained, and how effective would such a scheme be compared to centralized institutions?

To formalise this question,
we study a voluntary lottery in which authors accept a chance of random pre-review
rejection, reducing the reviewer burden and thereby reducing error in the review process.
When review quality is sufficiently sensitive to burden, the random loss of a few papers is more than compensated by the improved accuracy of decisions on all survivors.
The lottery is useful not as a policy prescription but because it isolates this tradeoff. 
It has a single free parameter (the rejection probability $L$), makes no quality judgments,
and poses the cooperation problem sharply: each scientist's decision to participate affects the noise level faced by all.
Unlike desk rejection quotas,
it requires no editorial discretion and no information beyond each author's willingness to enter.
Lottery mechanisms even have some precedent in science funding, where random allocation among shortlisted
proposals has been both proposed and piloted~\citep{fang2016research, gross2019contest}.

Our analysis proceeds as follows.
We first model the review process and show how the quality of accepted papers degrades as scale increases,
then introduce a pre-review lottery as a burden-reduction mechanism
and show that it improves the quality of published science (\cref{sec:cost_of_scale}).
We show that the optimal lottery takes the form of a quality threshold, with only low-quality papers entering the lottery (\cref{sec:lotteries}).
We then analyse the strategic equilibrium in which self-interested authors choose their own lottery participation
(\cref{sec:game_theory}), before discussing the implications of AI-driven submission growth (\cref{sec:ai}).
We show that even modest epistemic concern sustains voluntary participation, in turn,  narrowing the gap between self-interested behaviour and the social optimum.

\section{The cost of scale}
\label{sec:cost_of_scale}

We model a scientific venue (journal or conference) that receives $N$ submissions.
Each submission $i$ has true quality $q_i \in [0,1]$.
The editor observes a noisy signal of each paper's quality: the review score for paper $i$ is drawn from a truncated
normal distribution, with mean $q_i$ and standard deviation $\sigma$,
where $\sigma$ captures the noise inherent in the review process~\citep{zollman2023academic}.
The editor ranks submissions by score and accepts the top $\alpha N$ papers, yielding an acceptance rate $\alpha$.

We assume that review noise increases with the reviewer burden.
More papers per reviewer mean less time per assessment and a thinner pool of qualified referees,
both of which degrade the quality of evaluations.
We model this by tracking the fraction $\wbar$ of the $N$ submissions that receive full review.
The effective noise follows a power law:
\begin{equation}
  \label{eq:noise_scaling}
  \sigma_{\mathrm{eff}} = \sigma \cdot \wbar^{\,\beta}
\end{equation}
where $\sigma$ is the baseline noise (when all submissions are reviewed)
and $\beta > 0$ is the \emph{noise elasticity} --- the rate at which noise increases with load.
When $\wbar = 1$ (all papers reviewed), $\sigma_{\mathrm{eff}} = \sigma$;
any mechanism that reduces the fraction reaching review ($\wbar < 1$) also reduces noise.
The power-law form is consistent with empirical data and has some desirable modelling properties (see SI Section~S3 for details).

\begin{figure}[t]
  \centering
  \includegraphics[width=\textwidth]{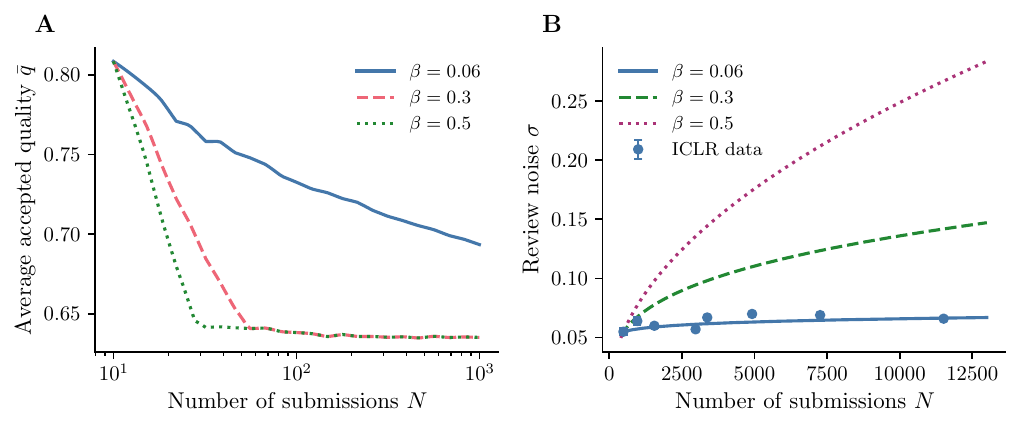}
  \caption{\textbf{Scale and the quality of published science.}
    \textbf{(A)}~Average accepted quality $\qbar$ as a function of venue size $N$ for three noise elasticities $\beta$,
    computed via Monte Carlo simulation ($M = 10{,}000$ replications; $\sigma = 0.3$, $\alpha = 10\%$).
    Higher $\beta$ produces steeper quality loss as venues grow.
    \textbf{(B)}~Empirical support for noise scaling: estimated review noise $\sigma$ from ICLR submission data (2017--2025; see SI Section~S3).}
  \label{fig:cost_of_scale}
\end{figure}

\Cref{fig:cost_of_scale} illustrates the central prediction: as the number of submissions grows,
review noise increases and the quality of accepted papers degrades.
This motivates the search for mechanisms that can break the link between scale and noise.
Various mechanisms already target this link (submission windows, reviewer mandates, excess-paper fees)
but remain ad hoc (see Discussion).

\label{sec:lottery_mechanism}
We study a simple mechanism: a lottery in which papers are randomly rejected before review with probability $L$,
surviving with probability $1 - L$.
The logic is straightforward: fewer surviving submissions reduce reviewer burden,
which translates into lower effective noise.
If all papers enter the lottery, a fraction $\wbar = (1 - L)$ survives to review, and the effective noise becomes $\sigma_{\mathrm{eff}} = \sigma \cdot (1 - L)^{\beta}$.
The lottery can only reduce noise, and does so more strongly when $\beta$ is large.
The tradeoff is that some papers, including potentially good ones, are lost before review.
Whether this tradeoff is worthwhile depends on how much noise the lottery eliminates relative to the fraction of good papers it sacrifices.

Note that the lottery does not change how many papers are published because the acceptance fraction $\alpha$ remains fixed: journals fill a fixed number of publication slots~\citep{card2020editors}, and acceptance rates at major venues are empirically stable even as submissions grow~\citep{berger_csconferences}.
The lottery only changes how accurately these papers are selected.
When the noise elasticity is large enough, even blanket adoption improves quality;
more generally, the lottery's value comes from its interaction with strategic behaviour.
We next ask who \emph{should} participate, for maximum benefit to science,
and then who \emph{will} participate when scientists act in their own interest.

\section{The optimal lottery}
\label{sec:lotteries}

We derive these results using a continuous approximation that yields tractable expressions for
acceptance probabilities, quality, and optimal participation.
The venue accepts a fixed fraction $\alpha$ of submissions, not a fixed number,
so both papers and acceptances scale with $N$;
thus, the limit converts a discrete ranking problem into a continuous threshold problem.

\label{sec:analytical}
Each author chooses a lottery participation level $p(q) \in [0,1]$,
and a submission with participation $p$ survives the lottery with probability $w = 1 - Lp$.
The aggregate survival fraction $\wbar = \int_0^1 w(q) f(q)\,dq$, where $f(q)$ is the quality density, determines the effective noise: $ \sigma_{\mathrm{eff}} = \sigma \cdot \wbar^{\,\beta}.$

The acceptance probability for a paper of quality $q$ is determined by the $(1-\alpha)$ quantile of the marginal
performance distribution (see SI Section~S2 for the full derivation).
The continuous approximation closely matches Monte Carlo simulations
(lines vs.\ markers in \cref{fig:planner_mc,fig:nash_prosociality}; systematic comparison in SI Section~S1).

\begin{figure}[t]
  \centering
  \includegraphics[width=\textwidth]{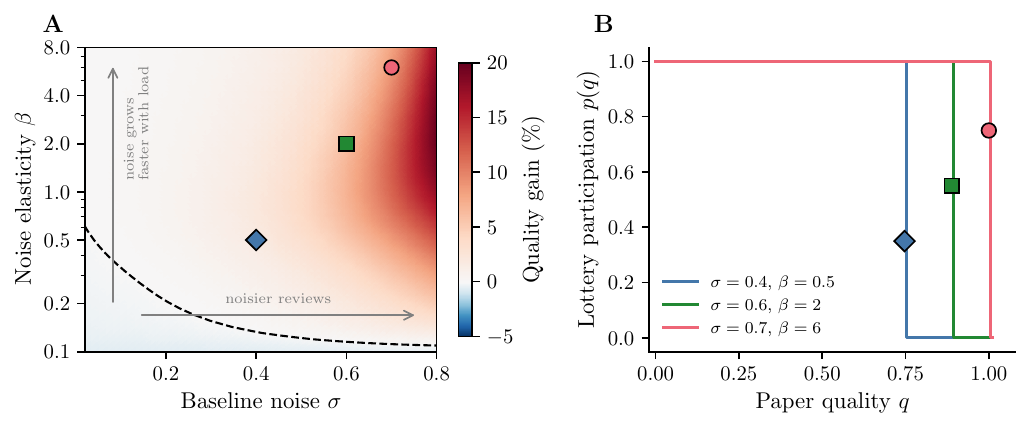}
  \caption{\textbf{Under full adoption, lotteries help when noise is sufficiently high; the optimal rule adapts to the regime.}
    \textbf{(A)}~Quality gain (colour) when all scientists enter the lottery, as a function of baseline noise $\sigma$
    and noise elasticity $\beta$, with acceptance rate $\alpha = 10\%$ and $L = 0.20$.
    The dashed contour marks the boundary where gain equals zero; above and to the right, the lottery improves quality.
    Markers indicate three representative parameter combinations shown in~(B).
    \textbf{(B)}~Socially optimal participation profiles $p(q)$ at the three marked points:
    deeper into the region where lotteries help, the optimal threshold rises and more papers enter the lottery.
    See SI Section~S3 for how the gain varies with acceptance rate.}
  \label{fig:phase_diagram}
\end{figure}

We first map the parameter space where full adoption of the lottery improves review quality.
\Cref{fig:phase_diagram} shows how the gain depends on baseline noise $\sigma$ and noise elasticity $\beta$.
Below the dashed contour, the lottery removes good papers without reducing noise enough to compensate;
above it, the noise reduction dominates and quality improves across a broad region.
Venues with higher noise or lower acceptance rates fall squarely in the regime where lotteries help,
though precise calibration remains difficult given the limited empirical data available (see SI Section~S3).

The lottery works best when participation is selective:
low-quality papers enter, reducing reviewer burden,
while high-quality papers do not enter, thus having maximal chance of being accepted.
The socially optimal rule (maximising the average quality of published science; since $\alpha$ is fixed, this is equivalent to maximising total quality) takes the form of a threshold:
all papers below a quality cutoff $\tau^*$ enter the lottery, while those above do not (\Cref{fig:phase_diagram}, panel B).
Low-quality papers contribute to reviewer burden but have little chance of acceptance; removing them reduces noise at almost no cost.
High-quality papers would likely be accepted, so removing them sacrifices real value (see SI Section~S2 for the proof).

\Cref{fig:phase_diagram}B shows how the optimal threshold adapts:
deeper into the region where lotteries help, more papers enter.
The quality improvement grows with noise (\cref{fig:phase_diagram}A),
confirming that the mechanism is most effective when review noise is sufficiently high.

\section{Voluntary participation}
\label{sec:game_theory}

We now ask whether scientists would voluntarily participate when acting in their own interest.
Each scientist with paper quality $q_i$ chooses a lottery participation probability $p_i \in [0,1]$.
The journal, too, is a strategic player: by putting in more effort,
it can lower the noise $\sigma$, but at an increasing marginal cost.
The lottery rejection probability $L$ is fixed by the venue exogenously
(see SI Section~S4 for alternative assumptions).

Each scientist's utility combines a private publication benefit, an epistemic term, and a rejection cost:
\begin{equation}
  \label{eq:payoff}
  U_i = b\,(1 - Lp_i)\,A(q_i)\,\qbar + s\,\qbar - c\,q_i\,\bigl[1 - (1-Lp_i)\,A(q_i)\bigr]
\end{equation}
where $A(q_i)$ is the acceptance probability for a paper of quality $q_i$,
$b$ is the private benefit of publication,
$s$ captures \emph{epistemic} motivation --- concern for the overall quality of published literature
regardless of one's own outcome ---
and $c$ is the cost of rejection, proportional to paper quality.
To tractably analyse this finite-$N$ game, we use a continuous approximation that treats the quality distribution as smooth but retains the finite impact of an individual's deviation on the aggregate reviewer burden (see SI Section~S2 for the derivation).

Since the payoff is linear in $(b, s, c)$, equilibria are invariant under rescaling $(b, s, c) \to (\lambda b,
\lambda s, \lambda c)$ for any $\lambda > 0$.
Only two quantities affect strategic behaviour: the ratio $r = b/(b+s)$, which captures the private--epistemic balance,
and the normalised cost $c/(b+s)$.
We hold the normalised cost fixed throughout and vary $r$: when $r = 1$, scientists are purely self-interested;
as $r$ decreases, they internalise the collective benefit of better review quality,
and the equilibrium shifts toward the social optimum.

Following~\citet{zollman2023academic},
the journal chooses the review noise $\sigma$ to maximise a payoff that trades off the quality of accepted papers
against the cost of thorough review:
\begin{equation}
  \Pi_J = \qbar - \wbar\,(1 + \sigma)^{-k}
\end{equation}
where $k > 0$ governs how steeply the cost of noise reduction increases as noise falls.
Because the lottery removes papers before review,
the total review cost scales with the surviving fraction $\wbar$.
The lottery thus operates through three channels: it shifts the quality composition of the reviewed pool,
it reduces noise by lightening the reviewer burden,
and it lowers the total cost of review.
The Nash equilibrium is a joint fixed point: scientists choose $p_i$ given the journal's noise level,
and the journal chooses $\sigma$ given the scientists' participation strategy.

\begin{figure}[t]
  \centering
  \includegraphics[width=\textwidth]{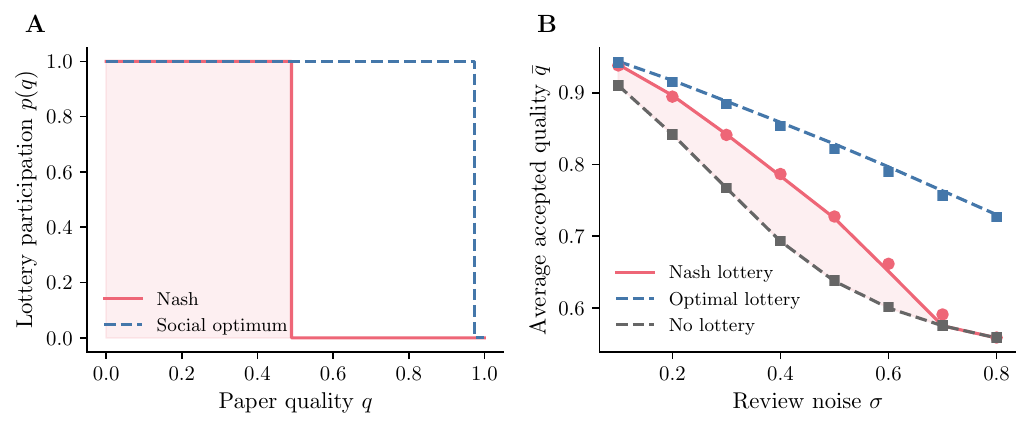}
  \caption{\textbf{Self-interest limits voluntary participation relative to the social optimum.}
    \textbf{(A)}~Participation profiles $p(q)$: the Nash equilibrium at $r = 0.33$ (solid) has a lower threshold
    than the social optimum (dashed); self-interested scientists under-participate.
    \textbf{(B)}~Average accepted quality $\qbar$ as a function of review noise under three regimes:
    the scientists' equilibrium lottery at $r = 0.33$ (solid red), the socially optimal lottery (dashed blue), and no lottery (dashed grey).
    Lines show the continuous approximation; markers show Monte Carlo simulation ($N = 100$).
    Each point shows the scientist-side equilibrium at a given noise level $\sigma$; varying $\sigma$ reveals how the gap between equilibrium and optimal participation grows with review noise.
    Parameters: $\beta = 8$, $L = 0.10$, $\alpha = 10\%$.}
  \label{fig:planner_mc}
\end{figure}

\Cref{fig:planner_mc} compares the Nash equilibrium to the social optimum.
Self-interested scientists under-participate: the threshold below which, in equilibrium, scientists submit to the lottery is lower than the social optimum,
and the quality gap between the two widens with noise (\cref{fig:planner_mc}B).
Yet even partial voluntary participation improves on the no-lottery baseline.
How much of the optimal gain is realised depends on the balance between private and epistemic motivation.

\begin{figure}[t]
  \centering
  \includegraphics[width=\textwidth]{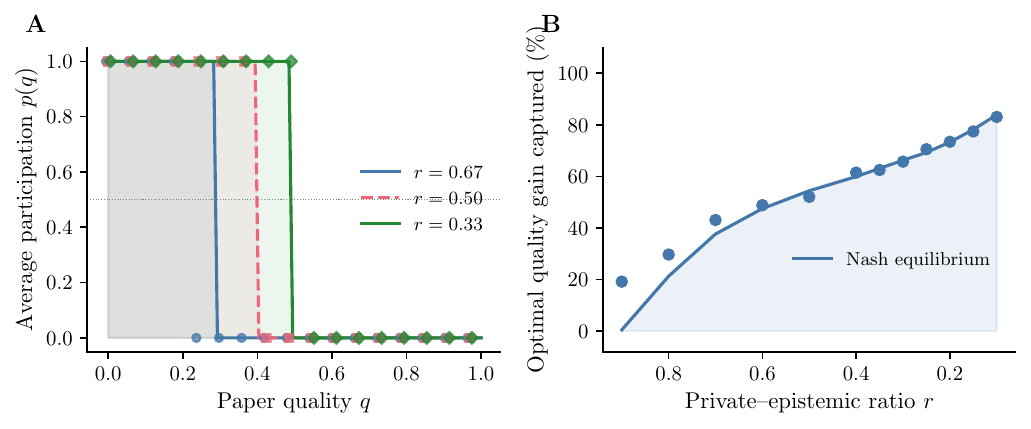}
  \caption{\textbf{Epistemic concern sustains voluntary participation and improves quality.}
    \textbf{(A)}~Equilibrium participation profiles $p(q)$ for three values of the private--epistemic ratio: $r = 0.67$ (two-thirds private), $r = 0.50$ (equal balance), and $r = 0.33$ (one-third private).
    Lines show the continuous approximation; dots show Monte Carlo simulation averages ($N = 100$).
    Noise is fixed at $\sigma = 0.3$.
    Lower $r$ (more epistemic) produces a higher participation threshold: more scientists enter the lottery when they weigh collective quality more heavily.
    \textbf{(B)}~Fraction of the optimal quality gain captured by the Nash equilibrium, as a function of $r$.
    As epistemic concern increases (lower $r$), the Nash equilibrium closes the gap to the social optimum, rising from near zero at $r = 0.9$ to more than 80\% at $r = 0.10$.
    Parameters: $\sigma = 0.3$, $\beta = 8$, $L = 0.10$, $\alpha = 10\%$, $c = 0.1$, $b = 1$.}
  \label{fig:nash_prosociality}
\end{figure}

\Cref{fig:nash_prosociality} shows that rational scientists sustain non-trivial lottery participation in equilibrium,
provided they are sufficiently epistemic in motivation.
The equilibrium has a clean structure: each scientist's best response is all-or-nothing,
and because the incentive to participate decreases with paper quality,
the Nash equilibrium is a threshold rule --- all papers below a cutoff $\tau^*$ enter, all above opt out
(see SI Section~S2 for the proof).
The threshold increases as the private--epistemic ratio $r$ decreases (i.e.,
as scientists place more weight on collective quality).
The gap between the social optimum and the Nash equilibrium narrows as $r$ decreases,
illustrating that epistemic concern for the quality of science is a partial substitute for centralised coordination.
Voluntary lotteries need no mandate, but they require scientists to see the quality of published science as partly their problem.
Monte Carlo simulations confirm that the continuous approximation is conservative: finite populations produce higher participation thresholds than the analytical prediction, so more papers enter the lottery than the theory predicts
(SI Table~S1), because discrete stochastic effects systematically favour voluntary participation.

\section{The AI scaling pressure}
\label{sec:ai}

Generative AI is lowering the cost of producing manuscripts faster than it is improving the capacity to evaluate them~\citep{noy2023experimental}.
As submission volume outpaces reviewer capacity, review noise $\sigma$ rises.
In our model, even a moderate increase in noise reduces accepted quality by 14\%;
a doubling of noise costs 22\%.
The lottery can recover much of this loss.
At the noise elasticity used throughout our analysis ($\beta = 8$),
the socially optimal lottery more than compensates for a doubling of noise,
restoring quality to above its pre-increase level.
The phase diagram (\cref{fig:phase_diagram}) maps the full picture:
the gain grows with both $\sigma$ and $\beta$,
meaning the worse the noise problem, the more the lottery helps.

These percentages may understate the damage at highly selective venues.
When a journal accepts 10\% of submissions, most decisions fall near the noise margin;
increased noise does not degrade all decisions equally but concentrates among borderline cases,
swapping papers that should have been accepted for papers that should not.

The evidence that submission growth is already outpacing review capacity is substantial.
Large-language-model access cuts professional writing time by roughly
40\%~\citep{noy2023experimental},
more than one-fifth of computer-science preprints on arXiv show
measurable evidence of large-language-model
modification~\citep{liang2024monitoring}, and at least 13\% of biomedical abstracts do the same~\citep{kobak2025delving}.
Submission volumes at major machine-learning venues have grown exponentially over the past decade,
and this growth has coincided with widespread perceptions of arbitrary editorial decisions ---
a pattern consistent with the quality degradation our model predicts.
The NeurIPS consistency experiments~\citep{price2014nips,
beygelzimer2023neurips} show that review noise was already substantial before AI-driven acceleration.
AI-driven submission growth is likely to push this further,
and AI-assisted research may amplify exploitation of well-explored problems at the expense of genuinely
novel inquiry~\citep{hao2026aitools},
compounding the signal-to-noise problem that reviewers face.

Higher noise pushes venues deeper into the region where the lottery helps,
increasing the potential gain from collective action.
These gains represent the ceiling on what any voluntary burden-reduction mechanism can achieve through coordinated participation.
How much of that ceiling is realised depends on the epistemic concern $r$
established in the previous section:
the more scientists internalise the collective benefit, the closer the outcome approaches the optimum.
Whether noise increases because of expanding research communities or because AI lowers production costs,
the underlying problem is the same: reviewer overload,
and the case for collective action strengthens as the gap between production and evaluation widens.

\section{Discussion}
\label{sec:discussion}

The voluntary lottery is deliberately simple ---
its value lies less in the specific mechanism than in what it reveals
about the potential for collective action in peer review.
Our central finding is that if scientists could collectively reduce the review burden ---
even through a mechanism as crude as random pre-review rejection ---
the resulting improvement in review quality would more than compensate for the papers lost,
provided the noise elasticity is sufficiently strong.
Our model does not capture the self-screening feedback posited by \citet{bergstrom2026screening},
whereby noisier review attracts more marginal submissions;
because this cycle is absent, the quality gains we report may be conservative.

In our model, the private--epistemic ratio $r$ determines how close voluntary participation comes to the social optimum:
epistemic concern substitutes for centralised coordination~\citep{fehr2002altruistic,heesen2021peerreview}.
The lottery threshold shares a structural parallel with the self-screening mechanism
of \citet{zollman2023academic}, where submission costs drive bottom-up self-selection.
Where Zollman et al. show that self-screening arises from imposed costs,
we show that a similar quality stratification can be sustained voluntarily through epistemic concern ---
without the welfare costs of imposed barriers.

This voluntary-versus-imposed distinction matters in practice.
Venues have already begun experimenting with burden-reduction mechanisms:
some philosophy journals restrict submissions to defined windows of the year~\citep{noauthor_journals_2026},
and computer science venues have introduced submission fees~\citep{ijcai2026_primary_paper}.
While more structural reforms could address the scale problem directly, they face large coordination challenges and resistance.
The lottery isolates this tradeoff and shows that imperfect adoption helps.
A venue need not enforce optimal participation to benefit.

Neither of the mechanism's requirements (sufficient noise elasticity $\beta$
and sufficient epistemic concern $r$) is a strong assumption.
Review noise is a universal feature of selective venues~\citep{dandrea2017editors, ellison2002evolving},
and behavioural evidence consistently shows that people contribute to collective goods
when they perceive the enterprise as worthwhile~\citep{fehr2002altruistic} ---
a description that fits most scientists' relationship to their field.
The analysis requires only that some scientists care, not that all do:
the SI shows the mechanism is robust to heterogeneous epistemic concern (SI Section~S4).

Because each scientist's influence on aggregate quality shrinks as the community grows,
community size matters.
This is not a limitation of the mechanism but a reflection of a structural constraint on collective action.
Peer review is a commons~\citep{hochberg2009tragedy}:
every scientist benefits from rigorous evaluation whether or not they contribute to providing it.
In small communities, reputation-based cooperation can sustain the commons,
because each member's behaviour is observable
and defection carries reputational cost~\citep{kandori1992social}.
As the community grows, monitoring becomes infeasible,
individual influence dilutes,
and the incentive to free-ride dominates.
Our model captures the dilution channel directly.
At $N = 100$, voluntary participation yields a meaningful quality gain; at $N = 500$, this roughly halves.
But even a modest increase in epistemic concern compensates: at $N = 500$, shifting from $r = 0.5$ to $r = 0.3$ roughly doubles the quality gain, recovering the level seen in smaller communities (SI Table~S1).
Without epistemic concern ($r \to 1$), the mechanism produces negligible gains regardless of community size.
The breakdown of reputation-based cooperation at scale reinforces this prediction in practice.

This pattern is characteristic of commons governance more broadly.
\citet{ostrom1990governing} identifies monitoring, graduated sanctions, and clearly defined boundaries
as design principles for sustainable collective management,
conditions that hold naturally in specialised journals and workshops
but break down at mega-conference scale,
where reviewers are drawn from an anonymous global pool
and no editor can track individual contributions.
Ostrom's eighth design principle (nested enterprises for larger-scale commons)
suggests the structural response:
federated review systems in which mega-venues decompose into community-scale tracks,
each small enough to restore the conditions under which voluntary cooperation is self-sustaining.
Some venues already approximate this through area-based review organisation;
our results suggest that community structure, not just reviewer incentives,
is key to sustaining review quality.

The mechanism's requirements (sufficient noise elasticity and epistemic concern) are most likely met where the problem is most acute.
\citet{beygelzimer2023neurips} show that making a venue more selective increases the arbitrariness of its decisions,
because the fraction of acceptances that fall within the review noise margin grows as the pool shrinks.
Our lottery is most effective in exactly this regime:
the venues where peer review is most arbitrary are those where voluntary collective action to
reduce noise would yield the greatest gains.
The peer review crisis, in other words, contains the seed of its own remedy: the worse the problem,
the stronger the incentive for scientists who care about the quality of published science to act.

\section*{Materials and Methods}
\label{sec:methods}

We model a scientific venue where scientists submit $N$ papers of quality $q_i \in [0,1]$,
an editor accepts the top fraction $\alpha$ based on noisy review scores,
and review noise scales with the fraction of papers reaching review
($\sigma_{\mathrm{eff}} = \sigma \cdot \wbar^{\,\beta}$; see Section~2 for full details).
Scientists may enter a voluntary pre-review lottery;
the journal chooses review effort to maximise its own payoff.
The Nash equilibrium is a joint fixed point of scientist and journal best responses.

The relevant strategies are \emph{threshold strategies} ---
step functions where all scientists with $q_i \leq \tau$ enter the lottery and those with $q_i > \tau$ do not.
In both the social optimum and the Nash equilibrium,
the threshold structure arises because the cost of losing a paper grows with quality
while the noise-reduction benefit does not (SI Section~S2).
Nash equilibria are identified by searching over all threshold strategies $\tau \in [0,1]$,
computing the journal's best-response noise level,
and verifying that no scientist can improve their payoff by unilateral deviation.
Monte Carlo simulations use $N = 100$ scientists and $M = 5{,}000$ replications per parameter combination.
The continuous approximation is evaluated on a 50-point quality grid via numerical integration.
The complete formal model specification is given in SI Section~S1; derivations of the acceptance probability,
lottery mechanism, payoff functions, and Nash equilibrium algorithm follow in SI Section~S2.

Source code is available at \url{https://github.com/juliangarcia/lottery-peer-review}.

\bibliographystyle{unsrtnat}
\bibliography{references}



\section*{Supplementary material}

# S1 Complete model specification

This section provides a detailed description of the model summarised in the main text. We specify the review process and noise-scaling mechanism, define the lottery and its interaction with review quality, introduce the payoff functions for scientists and the journal, and describe the Monte Carlo simulation used to validate the analytical results. We then report convergence checks and the $\varepsilon$-Nash characterisation of the equilibrium.

## S1.1 Review process

A scientific venue receives $N$ submissions, one from each scientist. Each scientist $i$ has a paper of true quality $q_i$ drawn independently from a uniform distribution on $[0, 1]$. The venue accepts a fixed fraction $\alpha$ of submissions.

To decide which papers to accept, an editor observes a noisy signal of each paper's quality, $X_i \sim \text{TruncatedNormal}(q_i, \sigma_{\text{eff}}^2, [0, 1])$, ranks all papers by their signal, and accepts the top $\alpha N$. The key feature of the review process is that noise increases with the reviewer burden. More papers per reviewer mean less time per assessment and a thinner pool of qualified referees. We model this as a power law: when a fraction $\bar{w}$ of submissions reach review, the effective noise is

$$\sigma_{\text{eff}} = \sigma \cdot \bar{w}^\beta \tag{S1}$$

where $\sigma$ is the baseline noise (when all submissions are reviewed) and $\beta > 0$ is the noise elasticity — the rate at which noise grows with the review burden. This functional form is empirically motivated by ICLR review data (see Section S3). When $\bar{w} = 1$ (all papers reviewed), $\sigma_{\text{eff}} = \sigma$; any mechanism that reduces the fraction reaching review ($\bar{w} < 1$) also reduces noise.

## S1.2 Lottery mechanism

We now introduce a pre-review lottery that can reduce the number of papers reaching the editor. Each scientist chooses a participation probability $p_i \in [0, 1]$.

A paper whose author chooses $p_i$ enters the lottery with probability $p_i$; if it enters, it is randomly rejected before review with probability $L \in (0, 1)$, surviving with probability $1 - L$. A paper that does not enter the lottery ($p_i = 0$) proceeds directly to review.

Each lottery entrant survives independently with probability $1 - L$, and each non-entrant survives with certainty. Because the lottery reduces the fraction of papers reaching review ($\bar{w} < 1$), it also reduces $\sigma_{\text{eff}}$, improving review quality for the papers that remain.

## S1.3 Payoffs and strategies

Each scientist with paper quality $q_i$ chooses $p_i$ to maximise their individual payoff:

$$U_i = b \cdot 1_{\text{published}} \cdot \bar{q} \; + \; s \cdot \bar{q} \; - \; c \cdot q_i \cdot (1 - 1_{\text{published}}) \tag{S2}$$

where $1_{\text{published}}$ indicates whether paper $i$ survives the lottery and is accepted by the editor, and $\bar{q}$ is the average quality of accepted papers.

The three terms in the scientists' utility capture distinct incentives:

- $b \geq 0$ weights the private benefit of publication, scaled by the quality of the venue;

- $s \geq 0$ captures *epistemic concern* — concern for the overall quality of published science, entering the payoff regardless of the scientist's own outcome;

- $c \geq 0$ is the rejection cost coefficient, scaled by paper quality $q_i$.

When $s = 0$, the scientist is purely self-interested; as $s$ grows, the scientist internalises the collective benefit of better review quality. The payoff is scale-invariant: only the private–epistemic ratio $r = b/(b+s)$ and the normalised cost $c/(b+s)$ affect equilibrium behaviour. Throughout the SI we work with the parameters $(b, s, c)$ directly; the correspondence is $r = 1/(1+s)$ when $b = 1$.

Each scientist's strategy is a choice of $p_i \in [0, 1]$. While this is formally a continuous choice, the equilibrium best response is binary: $p_i^* \in \{0, 1\}$ for all agents (Proposition 1). Furthermore, the set of agents who choose to enter ($p_i = 1$) always forms a lower interval $\{i : q_i \leq \tau\}$ for some threshold $\tau$ (Proposition 2).

Following Zollman et al. [1], the journal chooses the baseline review noise $\sigma$ to maximise:

$$\Pi_J = \bar{q} - \bar{w}(1 + \sigma)^{-k}, \tag{S3}$$

where $k > 0$ governs how steeply the cost of noise reduction increases as noise falls and $\bar{w}$ is the fraction of submissions surviving to review, so that the cost scales with the number of papers actually reviewed.

### S1.4 Equilibria and socially optimal lotteries

A Nash equilibrium is a joint fixed point: scientists choose $p_i$ given the journal's noise level, and the journal chooses $\sigma$ given the scientists' participation strategy. Details of the equilibrium algorithm we use to identify equilibria are given in Section S2.4.

The social optimum is defined by the threshold $\tau$ that maximises the average quality of accepted papers:

$$\tau^* = \arg\max_{\tau} \bar{q}(\tau), \tag{S4}$$

where $\bar{q}(\tau) = \mathbb{E}[\frac{1}{T}\sum_{j \in \text{accepted}} q_j]$ and $T$ is the number of accepted papers, estimated by Monte Carlo simulation over $M$ independent replications.

### S1.5 Monte Carlo simulation

For the finite stochastic game, we estimate payoffs via Monte Carlo simulation: for each strategy profile, $M$ independent replications of the review process are drawn, and expected payoffs are computed as sample averages. Nash equilibria are identified by searching over threshold strategies and verifying via deviation checks — for each candidate threshold, we check whether any agent gains from unilateral deviation. Best-response dynamics provide a complementary solution method: starting from an initial strategy profile, agents sequentially update to their best response until convergence.

### S1.6 Parameters

The model has three groups of exogenous parameters: venue characteristics ($\alpha$, $L$, $\beta$), payoff weights ($b$, $s$, $c$), and the journal's cost parameter $k$. The endogenous variables — each scientist's participation $p_i$, the journal's noise choice $\sigma$, and the resulting quality of accepted papers $\bar{q}$ — are determined in equilibrium. Unless otherwise stated, the following default values are used throughout:

| Parameter | Symbol | Default value |
| --- | --- | --- |
| Population size | $N$ | 100 |
| Acceptance rate | $\alpha$ | 10% |
| Baseline noise (reference value) | $\sigma$ | 0.3 |
| Noise elasticity | $\beta$ | 8 |
| Lottery rejection probability | $L$ | 0.10 |
| Journal review cost parameter | $k$ | 8 |
| Publication benefit | $b$ | 1 |
| Rejection cost | $c$ | 0.1 |
| Epistemic concern | $s$ | 1 |
| MC replications | $M$ | 5,000 |

We set $\beta = 8$ as an illustrative, high-elasticity parameter; while empirical data from a single conference venue suggests lower baseline elasticities in practice (see Section S3), using an amplified value clearly demonstrates the mechanism's structural dynamics and boundary conditions.

## S2 Analytical approximation

We use a continuous approximation that replaces the discrete ranking problem with a continuous threshold problem with tractable closed-form expressions for acceptance probabilities, quality, and payoffs. This section derives these expressions and proves the threshold structure of Nash equilibria. The continuous model provides structural insight (proving threshold existence, corner best responses, and monotonicity), while Monte Carlo simulation serves as quantitative ground truth.

### S2.1 Acceptance probability

The simulation in Section S1 uses a finite population of $N$ scientists. In this section we derive an analytical expression for the acceptance probability under the continuous approximation. The key insight is that the acceptance rate $\alpha$ does not vanish as $N$ grows, because the venue accepts a fixed *fraction* of submissions: both the number of papers and the number of acceptances scale with $N$.

Under the continuous approximation, the discrete ranking of $N$ papers is replaced by a continuous threshold problem in which each paper competes against a population distribution of perceived qualities rather than a finite set of competitors. Figure S1 confirms that this approximation is accurate even for moderate $N$.

Each paper has true quality $q$ drawn from $F(q)$ with density $f(q)$ on $[0, 1]$. The editor observes a noisy signal $X_q \sim \text{TruncatedNormal}(q, \sigma^2, [0, 1])$ with density:

$$g(x \mid q) = \frac{\phi\left(\frac{x-q}{\sigma}\right)}{\sigma\left[\Phi\left(\frac{1-q}{\sigma}\right) - \Phi\left(\frac{-q}{\sigma}\right)\right]} \quad (S5)$$

where $\phi$ and $\Phi$ are the standard normal PDF and CDF, respectively.

The perceived quality of a randomly drawn paper follows the mixture distribution with marginal density:

$$h(y) = \int_0^1 g(y \mid q) f(q) \, dq. \quad (S6)$$

The acceptance threshold $y^*$ is the $(1 - \alpha)$ quantile of this distribution:

$$\int_0^{y^*} h(y) \, dy = 1 - \alpha. \quad (S7)$$

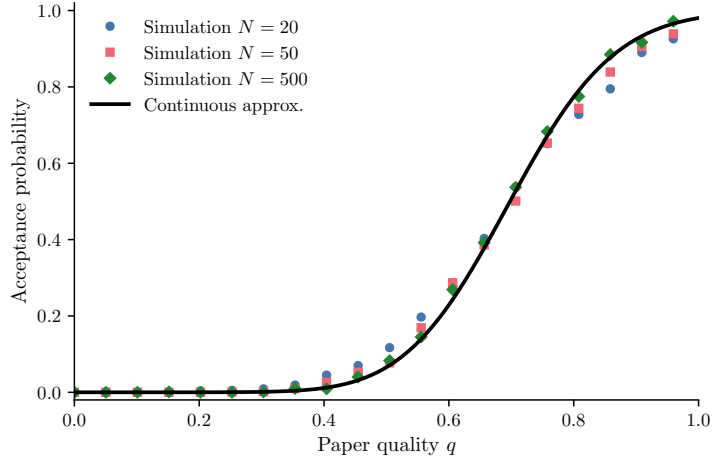

Figure S1: **Continuous approximation matches simulation.** Acceptance probability as a function of paper quality: Monte Carlo simulation (markers, $N = 20, 50, 500$) versus the continuous approximation (solid black line). Agreement improves as $N$ increases, confirming that the approximation is accurate for moderate populations.

A paper of true quality $q$ is accepted whenever its noisy signal exceeds this threshold, giving the acceptance probability:

$$P(\text{accept} \mid q) = \int_{y^*}^{1} g(x \mid q) \, dx = 1 - G(y^* \mid q) \tag{S8}$$

where $G(\cdot \mid q)$ is the CDF of TruncatedNormal$(q, \sigma^2, [0, 1])$.

In practice, we compute $h(y)$ by numerical integration over $q$, find $y^*$ by standard numerical root-finding on the quantile equation, and then evaluate $P(\text{accept} \mid q)$ for each quality level.

### S2.2 Special cases

Two limiting cases build intuition for the acceptance probability.

When noise vanishes ($\sigma \to 0$) and qualities are uniformly distributed ($f(q) = 1$), the editor observes true quality perfectly. Acceptance reduces to a deterministic cutoff: every paper above quality $1 - \alpha$ is accepted and every paper below is rejected:

$$y^* = 1 - \alpha, \tag{S9}$$

$$P(\text{accept} \mid q) = \begin{cases} 1 & \text{if } q > 1 - \alpha, \\ 0 & \text{if } q \leq 1 - \alpha. \end{cases} \tag{S10}$$

When noise is small but positive, this sharp step function softens into a smooth sigmoid. The threshold remains near $y^* \approx 1 - \alpha$, but papers close to the boundary now have intermediate acceptance probabilities:

$$P(\text{accept} \mid q) \approx 1 - \Phi\left(\frac{y^* - q}{\tilde{\sigma}}\right) \tag{S11}$$

where the effective width $\tilde{\sigma}$ accounts for truncation effects. Larger noise means a wider band of uncertainty around the cutoff.

### S2.3 Lottery mechanism derivations

Each submission survives the lottery with probability:

$$w(q) = 1 - L \, p(q) \tag{S12}$$

5

where $p(q) \in [0,1]$ is the participation level and $L \in [0,1]$ is the lottery rejection probability.

The average survivor fraction is:

$$\bar{w} = \int_0^1 w(q) f(q)\, dq. \tag{S13}$$

The overall lottery participation rate is:

$$\bar{p} = \int_0^1 p(q) f(q)\, dq. \tag{S14}$$

The effective noise decreases with the fraction of submissions removed by the lottery:

$$\sigma_{\text{eff}} = \sigma \cdot \bar{w}^\beta \tag{S15}$$

where $\beta > 0$ is the noise elasticity — the rate at which noise increases with load. Since $\bar{w} \leq 1$, the lottery can only reduce noise.

**Acceptance among survivors.** The lottery biases the reviewed pool. The acceptance rate among survivors is:

$$\alpha_{\text{sur}} = \frac{\alpha}{\bar{w}}. \tag{S16}$$

The performance density of a randomly selected survivor is:

$$h_{\text{sur}}(y) = \int_0^1 g(y \mid q) \frac{w(q) f(q)}{\bar{w}}\, dq. \tag{S17}$$

The acceptance threshold among survivors satisfies:

$$\int_0^{y^*} h_{\text{sur}}(y)\, dy = 1 - \alpha_{\text{sur}}. \tag{S18}$$

**Average quality of accepted papers.** The average quality of accepted papers is:

$$\bar{q} = \frac{\int_0^1 q\, w(q) f(q) [1 - G(y^* \mid q)]\, dq}{\int_0^1 w(q) f(q) [1 - G(y^* \mid q)]\, dq}. \tag{S19}$$

## S2.4 Payoff functions and Nash equilibrium

**Journal payoff and best response.** The journal's payoff trades off the quality of accepted papers against the cost of thorough review:

$$\Pi_J = \bar{q} - \bar{w}(1 + \sigma)^{-k} \tag{S20}$$

where $k > 0$ governs how steeply the cost of noise reduction increases as noise falls and the factor $\bar{w}$ reflects the per-paper nature of review costs. Higher $k$ makes thorough review more expensive, pushing the journal toward tolerating more noise at equilibrium.

The $\bar{w}$ factor in the cost term is not redundant with the noise-reduction channel $\sigma_{\text{eff}} = \sigma \cdot \bar{w}^\beta$. The two operate on different parts of the payoff: noise reduction is a demand-side effect (fewer papers improve each review), while cost scaling is a supply-side effect (fewer papers require less total review effort). Under this interpretation, the lottery benefits the journal through both channels simultaneously.

Given a scientist participation strategy $p(q)$, the journal's best response is the $\sigma$ that maximises $\Pi_J$.

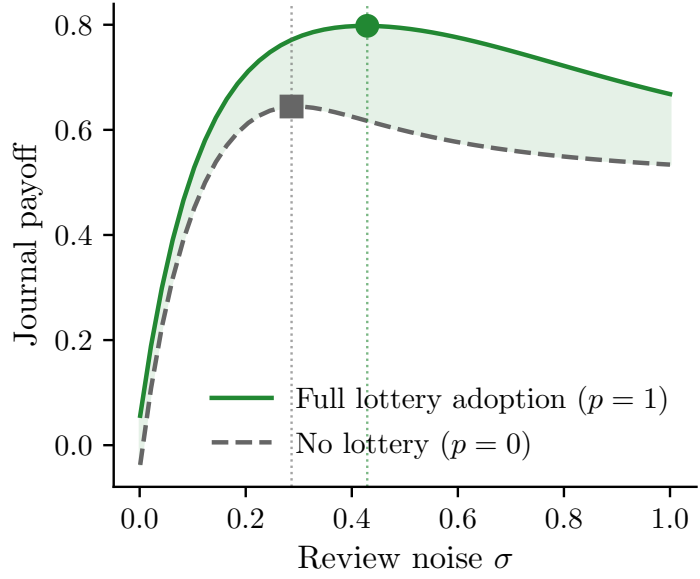

Figure S2: **The lottery raises the journal's peak payoff and shifts its optimum to higher noise.** Journal payoff as a function of review noise $\sigma$ under no lottery ($p=0$, dashed) and full lottery adoption ($p=1$, solid), with $k=8$, $\alpha=10\%$, $L=0.10$, $\beta=8$. Markers indicate the payoff-maximising noise level $\sigma^*$ under each regime. Because the lottery absorbs part of the submission load before review, the journal can tolerate a higher noise level at its optimum ($\sigma^* = 0.43$ vs. $0.29$) while achieving a higher peak payoff ($0.80$ vs. $0.64$). The shaded region shows the payoff gain from the lottery across the full noise range.

**Scientist payoff.** Replacing the indicator function in Equation (S2), with the lottery survival and acceptance probability gives:

$$U(q) = b\,(1-Lp)A(q)\,\bar{q} + s\,\bar{q} - c\,q\left[1-(1-Lp)A(q)\right] \tag{S21}$$

where $A(q) = 1 - G(y^* \mid q)$ is the acceptance probability conditional on surviving the lottery. The three terms are:

- $b\,(1-Lp)A(q)\,\bar{q}$ — publication benefit;
- $s\,\bar{q}$ — epistemic concern;
- $c\,q\,[1-(1-Lp)A(q)]$ — rejection cost.

**Deviation payoff.** To check whether a strategy profile $p(q)$ constitutes a Nash equilibrium, we compute the deviation payoff: what happens when a single scientist of quality $q$ unilaterally changes their participation from $p(q)$ to some alternative $p'$, while all other scientists maintain $p(\cdot)$. The deviator's change affects the aggregate survival fraction $\bar{w}$ and therefore $\sigma_{\text{eff}}$, the acceptance threshold, and $\bar{q}$. The strategy $p(q)$ is a best response if no unilateral deviation improves the scientist's payoff.

In a finite venue, opting out also creates a direct replacement effect by freeing an acceptance slot for a marginal paper. However, numerical analysis shows this direct effect is dominated by the indirect noise-reduction channel by a factor of 10 to 100, because it requires a paper to have both a high baseline acceptance probability and a quality far below the threshold; these conditions rarely overlap. For analytical clarity, our continuous approximation isolates the dominant noise-reduction channel.

7

**Social optimum.** The social optimum maximises $\bar{q}$, the average quality of accepted papers. The optimal policy is a threshold rule: papers with $q \leq \tau$ enter the lottery, while higher-quality papers do not. To see why, note that removing a paper of quality $q$ reduces noise (benefiting all survivors) but also removes its contribution to the accepted pool. The noise-reduction benefit is independent of which paper is removed, while the cost of removal increases in $q$, so the optimal rule removes papers in order of increasing quality, yielding a threshold.

The boundary case at $\tau = 1$ is illustrative. Here, full adoption is quality-blind, so noise falls to $\sigma(1-L)^\beta$ but the acceptance rate among survivors rises to $\alpha/(1-L)$. At $\sigma = 0$ the second effect dominates: $\bar{q}_{\text{full}} = 1 - \alpha/[2(1-L)] < 1 - \alpha/2$ — full lottery is worse than no lottery.

**Epistemic vs welfare social optimum.** The social optimum maximises published quality, rather than the welfare of scientists. Under threshold strategies and uniform quality on $[0,1]$, the two objectives are equivalent. To see this, note that aggregate welfare is

$$W = \int_0^1 U(q)\,dq = (b\alpha + s + c\alpha)\bar{q} - \tfrac{c}{2},$$

a strictly increasing affine function of $\bar{q}$, so maximising welfare and maximising quality select the same $\tau^*$ regardless of $b$, $s$, $c$, $\sigma$, or $\beta$.

The intuition is that the lottery reallocates quality across the accept/reject split, but cannot change either the total quality (fixed at $\int_0^1 q\,dq = 1/2$) or the accepted mass (fixed at $\alpha$ by venue capacity). Accepted-quality is therefore $\alpha\bar{q}$ and rejected-quality is $1/2 - \alpha\bar{q}$. Welfare combines three terms: a publication benefit on accepted quality ($b\alpha\bar{q}$), an epistemic gain in venue quality ($s\bar{q}$), and savings on rejection cost ($c\alpha\bar{q}$). All three terms grow with $\bar{q}$.

**Nash equilibrium algorithm.** The Nash equilibrium is a joint fixed point of two best-response mappings: scientists choose $p(q)$ given the journal's noise level $\sigma$, and the journal chooses $\sigma$ given $p(q)$. We solve this by searching over a grid of candidate $\sigma$ values and, for each, computing the scientists' fixed point.

**Scientists' fixed point.** Given $\sigma$, the scientists' equilibrium participation function $p^*(q)$ is computed by relaxation-based fixed-point iteration:

1. Initialise $p^{(0)}(q)$ on a discretised quality grid ($N$ points on $[0,1]$).

2. Compute the best response for each $q$ by grid search over $p' \in \{0, 1/M, 2/M, \ldots, 1\}$:

$$\text{BR}^{(n)}(q) = \arg\max_{p'} U(q; p', p^{(n)}_{-q}).$$

3. Update via squared-loss relaxation:

$$p^{(n+1)}(q) = p^{(n)}(q) - \eta \cdot 2\big(p^{(n)}(q) - \text{BR}^{(n)}(q)\big) \tag{S22}$$

where $\eta = 0.3$ is the learning rate, followed by clipping to $[0,1]$. By Proposition 1, the best response is a corner solution and the convex update preserves $[0,1]$, so in exact arithmetic the grid search and clipping are redundant. We retain both as numerical safeguards.

4. Repeat until $\max_q |p^{(n+1)}(q) - p^{(n)}(q)| < \delta$.

**Joint equilibrium.** For each candidate $\sigma$ on the grid, we compute the scientists' fixed point $p^*_\sigma(q)$, then check whether $\sigma$ is the journal's best response given $p^*_\sigma(q)$. A candidate is accepted as an equilibrium if the journal's best-response $\sigma^*$ satisfies $|\sigma^* - \sigma| < \delta_\sigma$ for a tolerance $\delta_\sigma = 10^{-6}$. Among all accepted candidates, we select the one with the highest journal payoff.

**Convergence.** In all parameter configurations tested, the relaxation scheme converges within 50 iterations to a final error below $10^{-3}$. The squared-loss convergence history is monotonically decreasing in typical runs.

## S2.5  Nash equilibria are threshold strategies

The socially optimal policy is provably a threshold rule (see "Social optimum" in Section S2). Here we show that Nash equilibria also take this form: all scientists with papers below some quality cutoff $\tau^*$ enter the lottery, and all scientists above $\tau^*$ do not. The argument has two steps. First, each scientist's best response is all-or-nothing: enter fully or not at all. Second, the incentive to enter the lottery decreases with paper quality, producing a clean threshold.

**Each scientist's decision is all-or-nothing**

Let $U_i(p_i)$ denote scientist $i$'s payoff as a function of their own participation, holding all other scientists' strategies fixed. We decompose $U_i(p_i) = D(p_i) + R(p_i)$, where $D$ captures the *direct effect* of participation on scientist $i$'s own payoff and $R$ the *indirect effect* mediated through the population aggregates. The direct effect is

$$D(p_i) = -L A(q_i)(b\bar{q} + c q_i) p_i + \text{const.},$$

where $A(q_i)$ is the acceptance probability conditional on passing the lottery stage and $\bar{q}$ is the average quality of accepted papers. $D$ is linear in $p_i$, so $D'$ is a constant and $D'' = 0$. The indirect effect $R$ enters the payoff through the population average $\bar{w} = \frac{1}{N}\sum_j (1 - L p_j)$, so $\partial \bar{w}/\partial p_i = -L/N$.

**Proposition 1** (Corner best response). *Whenever $(1+s) \ll N$, where $s$ is the epistemic-concern parameter, the direct effect dominates: $U_i$ is strictly monotone on $[0,1]$ and the best response satisfies $p_i^* \in \{0, 1\}$.*

*Sketch.* A linear function on $[0,1]$ has no interior maximum—it is maximised at one of the endpoints. We show that $U_i(p_i)$ is approximately linear for large $N$, so the best response is always a corner.

For any reference point $P \in [0,1]$, expand $U_i(p_i)$ around $P$:

$$U_i(p_i) = U_i(P) + U_i'(P)(p_i - P) + \tfrac{1}{2} U_i''(\xi)(p_i - P)^2,$$

for some $\xi$ between $P$ and $p_i$. The first two terms are linear in $p_i$, so the question reduces to whether the quadratic correction is negligible (i.e., whether $|U_i''|$ is small).

Since $D$ is linear in $p_i$, $D'' = 0$ exactly, so $U_i'' = R''$. The indirect effect $R$ depends on $p_i$ only through the population average $\bar{w}$, so we can write $R(p_i) = R(\bar{w}(p_i))$. Using the chain rule to find the second derivative we find:

$$R'' = \frac{\partial^2 R}{\partial \bar{w}^2}\left(\frac{\partial \bar{w}}{\partial p_i}\right)^2 + \frac{\partial R}{\partial \bar{w}}\frac{\partial^2 \bar{w}}{\partial p_i^2}.$$

The second term vanishes because $\bar{w}$ is linear in $p_i$. This leaves $R'' = (\partial^2 R/\partial \bar{w}^2)(\partial \bar{w}/\partial p_i)^2$. We bound each factor in turn. The first, $\partial^2 R/\partial \bar{w}^2$, measures how fast the indirect effect's slope changes with $\bar{w}$. The indirect channel enters the payoff through mean accepted quality $\bar{q}$, which carries combined weight $b + s = 1 + s$, so $\partial^2 R/\partial \bar{w}^2$ is bounded by a constant times $(1 + s)$. The second factor is $(\partial \bar{w}/\partial p_i)^2 = L^2/N^2$—the "one scientist among $N$" scaling. Multiplying:

$$|U_i''| = |R''| \leq C(1+s)\frac{L^2}{N^2}$$

for some constant $C$. This bound holds for all $p_i \in [0,1]$, so it holds at $\xi$ regardless of where $\xi$ falls. Since $|p_i - P| \leq 1$, the quadratic term is at most $\tfrac{1}{2} C(1+s) L^2/N^2$, which is negligible whenever $(1+s) \ll N$. The payoff is approximately linear, and the best response lies at a corner: $p_i^* \in \{0, 1\}$.

Since $U_i$ is approximately linear, its slope is approximately constant—we only need the sign of $U_i'(P) = D' + R'(P)$. When $A(q_i) > 0$, $D' < 0$ dominates and the payoff is decreasing: $p_i^* = 0$ (do not enter). When $A(q_i) \approx 0$, $D' \approx 0$ and $R' > 0$ dominates: $p_i^* = 1$ (enter). At the threshold quality $\tau^*$, the two payoffs are equal by continuity, and the threshold is determined by where $\Delta(q)$ crosses zero (Proposition 2). $\square$

**The incentive to enter decreases with paper quality**

By Proposition 1, every scientist plays either $p = 1$ or $p = 0$. Define the net incentive to enter for a scientist with paper quality $q$ as

$$\Delta(q) = U(q, p = 1, \mathrm{p}_{-q}) - U(q, p = 0, \mathrm{p}_{-q}).$$

**Proposition 2** (Unique threshold). *Under the conditions of Proposition 1, $\Delta(q)$ is continuous and strictly decreasing in $q$, with $\Delta(0) > 0$ and $\Delta(1) < 0$. Consequently there is a unique threshold $\tau^* \in (0,1)$ such that all scientists with $q < \tau^*$ enter the lottery and all with $q > \tau^*$ do not.*

*Sketch.* We show (i) $\Delta$ is strictly decreasing on $[0,1]$ and (ii) $\Delta(0) > 0$, $\Delta(1) < 0$; continuity then gives a unique zero crossing.

*(i) Strict decrease.* The dominant term of $\Delta$ is the direct personal cost of entering,

$$\Delta_{\text{direct}}(q) = -L A(q)(b\bar{q} + cq),$$

whose derivative is

$$\frac{d\Delta_{\text{direct}}}{dq} = -L\bigl[A'(q)(b\bar{q} + cq) + A(q)c\bigr] < 0,$$

strictly negative and of order one (independent of $N$). Two mechanisms drive the decrease: better papers are more likely to be accepted ($A'(q) > 0$), so they lose more from random pre-review rejection; and the rejection cost $cq$ grows with paper quality.

By the same chain-rule argument used in Proposition 1, aggregate shifts in $(\bar{w}, \sigma_{\text{eff}}, y^*, \bar{q})$ contribute corrections to $d\Delta/dq$ of size $O((1+s)/N)$, small relative to the order-one dominant term at the paper's parameters. Hence $d\Delta/dq < 0$ throughout $[0,1]$.

*(ii) Boundary values.* At $q = 0$, the acceptance probability $A(0)$ is essentially zero under any realistic noise level, so the direct entry cost vanishes. The aggregate benefit remains positive: a lowest-quality paper that enters the lottery and is randomly rejected reduces the noise in the review pool and slightly improves mean accepted quality. Hence $\Delta(0) > 0$. At $q = 1$, the direct cost $LA(1)(b\bar{q}+c)$ is substantial — a top-quality paper is near-certain to be accepted without the lottery — and dominates the aggregate benefit of size $O(s/N)$. Hence $\Delta(1) < 0$.

$\Delta$ is continuous, strictly decreasing, positive at $q = 0$ and negative at $q = 1$, so it crosses zero exactly once at some $\tau^* \in (0,1)$. $\square$

**Robustness to the cost specification.** The threshold structure does not depend on the quality-proportional cost $cq$. With a flat rejection cost $c$ (independent of $q$), the direct term becomes $\Delta_{\text{direct}}(q) = -LA(q)(b\bar{q} + c)$, whose derivative is $-LA'(q)(b\bar{q}+c) < 0$ — still strictly decreasing. The $A(q)$ channel alone suffices for monotonicity; the quality-proportional cost strengthens the result but is not required.

**Threshold persistence across population sizes**

A natural concern about Propositions 1 and 2 is whether the equilibrium threshold survives at larger populations. As $N$ grows, each scientist's individual influence on the aggregate is of order $1/N$, which might suggest that the incentive to enter vanishes and the equilibrium collapses to $\tau^* = 0$.

We test this directly by running the Nash equilibrium search at $N \in \{20, 50, 100, 200, 500\}$. Table S1 reports the Monte Carlo equilibrium threshold, the analytical prediction, and the quality gain over the no-lottery baseline. The threshold decreases with $N$ but remains substantial across the tested range, and the quality gain stays positive at every population size.

| $N$ | $\tau^*_{\text{MC}}$ | $\tau^*_{\text{cont}}$ | $\Delta \bar{q}_{\text{MC}}$ |
|---:|---:|---:|---:|
| 20  | 0.63 | 0.53 | 0.045 |
| 50  | 0.51 | 0.41 | 0.040 |
| 100 | 0.46 | 0.32 | 0.030 |
| 200 | 0.33 | 0.24 | 0.024 |
| 500 | 0.31 | 0.10 | 0.016 |

Table S1: **Threshold persistence across population sizes.** Monte Carlo equilibrium threshold $\tau^*_{\text{MC}}$, continuous approximation $\tau^*_{\text{cont}}$, and quality gain $\Delta \bar{q}_{\text{MC}}$ over the no-lottery baseline. Parameters: $\sigma = 0.3$, $\beta = 8$, $L = 0.05$, $\alpha = 10\%$, $s = 1$, $c = 0.1$, $M = 5{,}000$.

Two features of the data deserve comment. First, the MC threshold sits consistently above the continuous approximation, and the gap widens with $N$: at $N = 500$, $\tau^*_{\text{MC}} \approx 0.31$ while $\tau^*_{\text{cont}} \approx 0.10$. The continuous approximation is therefore a conservative lower bound on cooperation: by evaluating payoffs at the expected aggregate composition rather than averaging over discrete stochastic realisations, it smooths away the variance that systematically boosts the epistemic payoff of volunteering (a Jensen's-inequality effect that favours participation in finite populations). Second, the quality gain $\Delta \bar{q}$ also shrinks with $N$, from $\sim 0.045$ at $N = 20$ to $\sim 0.016$ at $N = 500$. The mechanism therefore retains its qualitative character at large populations but becomes quantitatively weaker as individual influence dilutes.

We also verified threshold structure empirically across motivation regimes: unrestricted best-response dynamics from 36 random starting conditions ($N = 100$, three private–epistemic ratios) all converge to threshold-shaped profiles with zero ordering violations.

### S2.6 Validation of the analytical approximation

The continuous approximation shows close agreement with finite-population Monte Carlo simulation across the full noise range, for both the Nash equilibrium and the social optimum (Figure S3).

To further characterise the accuracy of the continuous approximation at finite $N$, Table S2 reports the optimal threshold $\tau^*$ and quality gain $\Delta \bar{q}$ across several population sizes. Both converge as $N$ increases, and the analytical predictions are accurate even at moderate $N$.

**$\varepsilon$-Nash characterisation.** The analytical threshold $\tau^*$ is not an exact Nash equilibrium of the finite stochastic game (the maximum deviation gain is positive under Monte Carlo estimation), but it is an $\varepsilon$-Nash equilibrium [2] with $\varepsilon \approx 0.003$. This means no agent can gain more than 0.3% in expected payoff by unilateral deviation.

## S3 Empirical calibration of noise elasticity

The power-law noise model $\sigma_{\text{eff}} = \sigma \cdot \bar{w}^\beta$ requires two empirical inputs: the baseline noise $\sigma$ and the noise elasticity $\beta$. Calibrating either number against real venues is hard because peer-review scores are almost never released: most journals and conferences treat reviewer ratings as confidential, and the published experimental studies [3, 4] cover only a single venue-year with no variation in $N$. Machine-learning conferences on OpenReview are a rare exception, and ICLR in particular has over a decade of fully anonymised scores spanning a wide range of conference sizes, making it the best available setting for calibrating a size-dependent noise model. The numbers reported below should

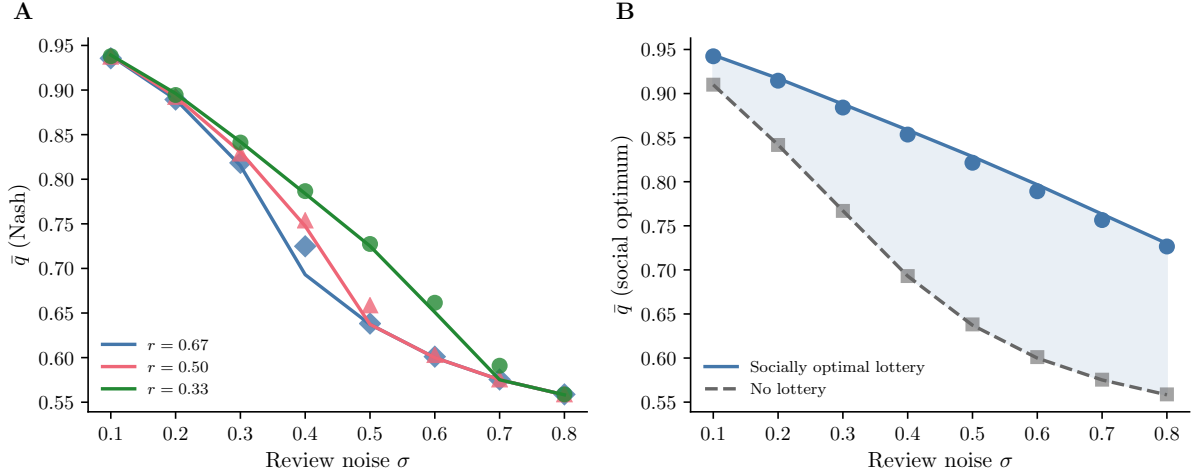

Figure S3: **Continuous approximation matches finite-population Monte Carlo simulation.** Average quality of accepted papers $\bar{q}$ as a function of review noise $\sigma$. Lines show the continuous approximation; markers show Monte Carlo simulation ($N = 100$, $M = 5{,}000$). **(A)** Nash equilibrium for three private–epistemic ratios ($r = 0.67, 0.50, 0.33$, corresponding to $s = 0.5, 1, 2$). Higher epistemic concern (lower $r$) sustains larger quality gains across all noise levels; the gains narrow and disappear beyond $\sigma \approx 0.5$–$0.7$, where the equilibrium collapses to no participation. **(B)** Social optimum and the no-lottery baseline.

|   | Continuous approx. | | | Monte Carlo | | |
|---|---|---|---|---|---|---|
| $N$ | $\tau^*$ | $\bar{q}$ | $\Delta\bar{q}$ | $\tau^*$ | $\bar{q}$ | $\Delta\bar{q}$ |
| 20 | 0.947 | 0.8578 | 0.0747 | 1.000 | 0.8448 | 0.0601 |
| 30 | 0.966 | 0.8510 | 0.0746 | 1.000 | 0.8406 | 0.0659 |
| 50 | 0.980 | 0.8455 | 0.0744 | 0.959 | 0.8394 | 0.0698 |
| 75 | 0.986 | 0.8427 | 0.0743 | 1.000 | 0.8396 | 0.0691 |
| 100 | 0.990 | 0.8413 | 0.0742 | 0.980 | 0.8380 | 0.0715 |

Table S2: **Grid convergence: optimal threshold and quality gain.** Continuous approximation versus Monte Carlo simulation for several population sizes $N$. Parameters: $\sigma = 0.3$, $\beta = 8$, $L = 0.10$, $\alpha = 10\%$, $M = 5{,}000$.

therefore be read as illustrative of what the mechanism looks like at one unusually open venue, not as venue-agnostic constants.

### S3.1 Dataset and estimation of $\sigma$

**Data source.** We use anonymised review scores from ICLR submissions (2017–2025), obtained from the ICLR dataset of González-Márquez and Kobak [5], a complete scrape of the OpenReview platform.[1] ICLR uses an ordinal 1–10 score and assigns at least three reviewers per paper; the mean number of reviews per paper rises from 3.0 in 2017 to 4.1 in 2025 (Table S3). After restricting to papers with $n \geq 2$ scores, the panel contains roughly 35,000 papers across nine conference years.

**Mapping reviewer disagreement onto the model.** The model in Section S1.1 treats $\sigma$ as the standard deviation of the editor's *aggregate* quality signal $X_i$, not of any individual reviewer. We therefore estimate $\sigma$ in three steps:

---
[1]Data and processing pipeline at https://github.com/berenslab/iclr-dataset.

| Year | $N$ | $\bar{n}$ | $\hat{\sigma}_{\text{rev}}$ | $\hat{\sigma}$ | 95% CI |
|------|-----|-----|-------|-------|----------------|
| 2017 | 489 | 3.1 | 0.096 | 0.055 | [0.052, 0.058] |
| 2018 | 967 | 3.0 | 0.111 | 0.064 | [0.061, 0.067] |
| 2019 | 1,554 | 3.0 | 0.104 | 0.060 | [0.058, 0.061] |
| 2020 | 2,561 | 3.0 | 0.155 | 0.089 | [0.087, 0.092] |
| 2021 | 2,974 | 3.9 | 0.112 | 0.057 | [0.056, 0.058] |
| 2022 | 3,375 | 3.9 | 0.131 | 0.067 | [0.066, 0.068] |
| 2023 | 4,919 | 3.8 | 0.136 | 0.070 | [0.069, 0.072] |
| 2024 | 7,261 | 3.9 | 0.134 | 0.069 | [0.068, 0.070] |
| 2025 | 11,513 | 4.1 | 0.132 | 0.066 | [0.065, 0.067] |

Table S3: **Per-year estimates of review noise from ICLR data.** $N$ is the number of papers with $n \geq 2$ scores; $\bar{n}$ is the mean number of reviews per paper; $\hat{\sigma}_{\text{rev}}$ is the average within-paper standard deviation on the rescaled $[0,1]$ scale; $\hat{\sigma} = \hat{\sigma}_{\text{rev}}/\sqrt{\bar{n}}$ is the implied noise of the aggregate review signal. The 2020 conference (greyed in figures) is excluded from subsequent fits as a COVID-era outlier: a sudden expansion of the reviewer pool and the rapid switch to remote review elevated within-paper disagreement well above trend.

1. **Rescale** each raw score to the model's quality scale, $\tilde{s}_{p,r} = (s_{p,r} - 1)/9 \in [0, 1]$, where $s_{p,r}$ is reviewer $r$'s score for paper $p$.

2. **Within-paper noise.** For each paper with $n_p \geq 2$ reviews, compute the sample standard deviation $\hat{\sigma}_{\text{rev},p} = \text{sd}(\tilde{s}_{p,1}, \ldots, \tilde{s}_{p,n_p})$. This estimates the noise of a single reviewer's score around the paper's true quality.

3. **Aggregate-signal noise.** Convert to the noise of the mean of $n_p$ reviews, $\hat{\sigma}_p = \hat{\sigma}_{\text{rev},p}/\sqrt{n_p}$ (the usual averaging argument: averaging $n$ independent noisy scores produces a mean with $\sqrt{n}$ times less noise). This is the standard error the editor would face when ranking by the mean review score, and it is the quantity that maps onto the model's $\sigma$.

Concretely, for 2025 the year-level averages are $\hat{\sigma}_{\text{rev},y} = 0.132$ and $\bar{n}_y = 4.1$, giving $\hat{\sigma}_y = 0.132/\sqrt{4.1} \approx 0.065$ (Table S3). Because ICLR reviews every submission ($\bar{w} = 1$), the resulting estimate is of the *baseline* noise $\sigma$ in $\sigma_{\text{eff}} = \sigma \bar{w}^\beta$, not of some post-intervention $\sigma_{\text{eff}}$. We then average the per-paper $\hat{\sigma}_p$ across papers within each year to obtain the annual estimate $\hat{\sigma}_y$, reporting a bootstrap 95% confidence interval (2,000 resamples); averaging (rather than pooling all scores) keeps within-paper reviewer disagreement separated from between-paper quality variation. Averaging $\hat{\sigma}_y$ across years (excluding 2020) yields the headline calibration $\sigma \approx 0.065$. Year-level estimates are reported in Table S3 and lie in the narrow range $[0.055, 0.070]$ once 2020 is excluded.

**Statistical modelling assumptions.** The estimator rests on four standard assumptions: (i) reviewers' rescaled scores are conditionally independent noisy estimates of paper quality; (ii) the ordinal 1–10 scale can be treated as cardinal; (iii) noise is homoscedastic across the quality range within a year; and (iv) reviewer-paper assignment is exogenous. (i)–(iii) are textbook measurement assumptions; the within-paper reviewer disagreement we measure is in line with prior reports for ML venues [3, 4]. (iv) is mildly violated by ICLR's bidding system, which probably biases $\hat{\sigma}$ downward (expert reviewers disagree less), making our calibration a lower bound on the noise a generic editor would face.

### S3.2 Functional form and calibrated $\beta$

To estimate $\beta$ we fit four candidate forms relating the conference size $N$ to the year-level noise $\sigma(N)$, where $a$, $\hat{b}$, $N_0$, and $\sigma_{\max}$ are fitted constants:

13

1. **Power law:** $\sigma(N) = a \cdot N^{\hat{b}}$
2. **Logarithmic:** $\sigma(N) = a + \hat{b} \log N$
3. **Saturating:** $\sigma(N) = \sigma_{\max}(1 - e^{-N/N_0})$
4. **Linear:** $\sigma(N) = a + \hat{b} N$

By AIC, the power-law and logarithmic forms are nearly tied and both substantially outperform the saturating and linear alternatives. The data alone therefore do not select between the two top-ranked forms; the choice has to be made on structural grounds.

**Why we adopt the power law.** The power-law and logarithmic forms fit the data equally well (tied by AIC), so the choice has to be made on what the parameter $\beta$ *means* across venues. Only the power law makes the lottery's relative effect a function of the reduction rate $\bar{w}$ alone:

$$\frac{\sigma(N\bar{w})}{\sigma(N)} = \frac{a(N\bar{w})^{\hat{b}}}{aN^{\hat{b}}} = \bar{w}^{\hat{b}}, \tag{S23}$$

the same fractional drop at every venue, regardless of raw size $N$ or baseline $\sigma$. The logarithmic form gives $\sigma(N\bar{w}) - \sigma(N) = \hat{b} \log \bar{w}$: an absolute reduction identical for a venue with $\sigma = 0.50$ and one with $\sigma = 0.05$, which is implausible. A calibration under the power law therefore transports across venues without recalibration; under the log form, $\hat{b}$ would be a property of the specific venue we fit on.

**Calibrated estimate.** Taking logs of $\sigma(N) = aN^{\beta}$ gives $\log \sigma = \log a + \beta \log N$, a straight line in log–log coordinates whose slope is $\beta$. We therefore estimate $\beta$ by ordinary least squares of $\log \hat{\sigma}_y$ on $\log N_y$, which is consistent with the multiplicative scaling argued for above. The fit yields $\hat{\beta} \approx 0.06$ ($R^2 = 0.50$, AIC $= -62.7$). Including 2020 leaves $\hat{\beta}$ essentially unchanged at 0.058 but collapses $R^2$ to 0.16, reflecting the atypical within-paper disagreement in that COVID-era conference; the point estimate is thus robust to the exclusion. The fit implies a noise-reduction factor $0.90^{0.06} \approx 0.99$ for a 10% reduction in the reviewed pool. For each $(\sigma, \alpha)$ pair, the gain from a full-participation lottery crosses zero at a critical value $\beta^*$; Table S4 reports $\beta^*$ across several combinations. At the ICLR calibration ($\sigma = 0.065$, $\alpha = 0.10$), $\beta^* \approx 0.99$; the threshold falls sharply with baseline noise, reaching $\beta^* \approx 0.08$ at $\sigma = 0.5$. Across the $\sigma$ range shown, $\beta^* > \hat{\beta}$, but the gap narrows from sixteenfold at ICLR to within a factor of two at high noise. A venue's position relative to the feasibility boundary therefore depends on $\sigma$ as much as on $\beta$ — the joint dependence that the phase diagram in the main text visualises directly.

**Interpretation.** $\hat{\beta}$ is the slope of aggregate review noise against submission volume *across* ICLR years — a descriptive quantity that tells us how noise changed as the venue grew. The lottery mechanism responds to a different quantity: the within-year causal effect of reducing load on a fixed reviewer pool. These differ because the mean number of reviews per paper rose from 3.0 in 2017 to 4.1 in 2025 (Table S3), which mechanically dampens the across-year slope through the $1/\sqrt{n}$ factor in $\hat{\sigma}$. Other changes at the conference over this period may push in the same direction, but we cannot document them from data alone. The headline $\hat{\beta} \approx 0.06$ should therefore be read as a lower bound on the model's $\beta$: the calibration confirms the direction of the load-sensitivity channel (review noise scales with reviewer burden) without pinning down its magnitude, and reflects a single conference venue rather than a universal calibration target. The mechanism requires $\beta$ above approximately 1 to activate (Table S4); our illustrative $\beta = 8$ is well above this threshold. Because the model functions as an illustrative thought experiment rather than a literal policy projection, our strategic simulations use this high-elasticity parameter to isolate the theoretical mechanics of the pre-review lottery, making the system's phase transitions and collective action trade-offs structurally visible.

14

| $\sigma$ | Acceptance rate $\alpha$ | $\beta^*$ (critical) |
|---|---|---|
| 0.065 | 10% | 0.99 |
| 0.065 | 20% | >20 |
| 0.065 | 32% | >20 |
| 0.1 | 10% | 0.48 |
| 0.2 | 10% | 0.20 |
| 0.3 | 10% | 0.12 |
| 0.3 | 32% | 0.49 |
| 0.5 | 10% | 0.08 |

Table S4: **Critical noise elasticity.** Minimum $\beta$ for which the full-participation lottery ($L = 0.10$) produces positive quality gains over the no-lottery baseline, across representative combinations of baseline noise $\sigma$ and acceptance rate $\alpha$. The threshold falls rapidly with both $\sigma$ and selectivity; at moderate noise ($\sigma \geq 0.2$) and low acceptance rates, small $\beta$ suffice.

Table S5: **Lottery participation decreases as scientists become more privately oriented.** Nash equilibrium as a function of the private–epistemic ratio $r = b/(b+s)$, with $c = 0.1$ held fixed. Other parameters: $N = 50$, $\sigma = 0.3$, $\beta = 8$, $L = 0.10$, $\alpha = 10\%$.

| $r = b/(b+s)$ | $\tau^*$ | $n_{\text{in}}$ | $\bar{q}_{\text{Nash}}$ |
|---|---|---|---|
| 0.1 | 0.71 | 37 | 0.823 |
| 0.2 | 0.63 | 33 | 0.819 |
| 0.3 | 0.59 | 28 | 0.814 |
| 0.5 | 0.39 | 22 | 0.807 |
| 0.7 | 0.37 | 18 | 0.798 |
| 0.9 | 0.14 | 7 | 0.787 |

# S4 Extensions and robustness

We examine the robustness of our results along three dimensions: sensitivity to the private–epistemic balance, heterogeneity in epistemic concern across scientists, and the moral hazard that arises when the journal controls the lottery intensity.

## S4.1 Sensitivity to the private–epistemic ratio

The payoff function (Equation 4 in the main text) is linear in $(b, s, c)$, so equilibria are invariant under rescaling $(b, s, c) \to (\lambda b, \lambda s, \lambda c)$ for any $\lambda > 0$. Only the ratio $r = b/(b+s)$ and the normalised cost $c/(b+s)$ affect strategic behaviour. Setting $b = s = 1$ (i.e. $r = 0.5$) is a convenience; here we verify that the qualitative results hold across the full range of private–epistemic balance.

Table S5 reports the Nash equilibrium threshold $\tau^*$, the number of participating scientists $n_{\text{in}}$, and the equilibrium quality $\bar{q}$ as a function of $r$, with $c = 0.1$ held fixed. The qualitative result — that voluntary lottery participation is sustained in equilibrium and improves quality — holds for all $r < 1$. As $r$ increases (scientists weight private benefit more heavily), the threshold drops, fewer scientists enter, and the quality gain over the no-lottery baseline diminishes. The quantitative gap between the Nash equilibrium and the social optimum widens as scientists become more privately oriented.

The normalised rejection cost $c/(b+s)$ has a modest, monotonic effect: higher cost lowers the equilibrium threshold (fewer papers enter the lottery) but does not qualitatively change the dependence on $r$. Across a 50-fold range of $c/(b+s)$ (from 0.01 to 0.50), the fraction of the planner's gain captured by the Nash equilibrium varies by roughly 5–10 percentage points within a given $r$, compared to a 30 percentage-point spread across $r$ values.

Table S6: **Robustness to the quality distribution.** Nash equilibrium threshold, participation, and quality under uniform and Beta(2,5) quality distributions. All equilibria are threshold-shaped with zero violations. Parameters: $\sigma = 0.3$, $\beta = 8$, $L = 0.10$, $\alpha = 10\%$, $N = 100$.

| Distribution | $s$ | $n_{\text{in}}$ | $\bar{q}_{\text{Nash}}$ | Threshold? |
|---|---|---|---|---|
| Uniform | 0.5 | 27 | 0.790 | ✓ |
| Uniform | 1.0 | 36 | 0.797 | ✓ |
| Uniform | 2.0 | 46 | 0.805 | ✓ |
| Beta(2,5) | 0.5 | 10 | 0.417 | ✓ |
| Beta(2,5) | 1.0 | 29 | 0.431 | ✓ |
| Beta(2,5) | 2.0 | 54 | 0.451 | ✓ |

### S4.2 Robustness to the quality distribution

The main analysis assumes paper qualities are uniformly distributed on $[0, 1]$. To test robustness, we repeat the social-optimum and Nash equilibrium analyses with a right-skewed Beta(2,5) distribution, which concentrates mass on lower-quality papers (mean $\approx 0.29$) and has fewer high-quality papers — arguably a more realistic shape.

Table S6 reports the results. The threshold equilibrium structure survives intact: all six Nash equilibria under Beta(2,5) are threshold-shaped with zero ordering violations. The qualitative pattern — more epistemic concern leads to more participation and higher quality — is identical across both distributions. The social optimum also yields a threshold under Beta(2,5) ($\tau^* = 0.614$, compared to $\tau^* = 1.000$ under uniform), with a similar quality gain.

### S4.3 Robustness to heterogeneous epistemic concern

The main analysis assumes that all scientists share the same epistemic concern parameter $s$. In practice, scientists differ in how much weight they place on the collective quality of published science. A natural worry is that heterogeneity could unravel the mechanism via a weakest-link effect: if a substantial fraction of scientists are purely self-interested ($s = 0$), they might free-ride on the epistemic concern of others.

To test this, we allow $s_i$ to vary across scientists while holding the population mean fixed at $\bar{s} = 1$. We consider two illustrative scenarios alongside the homogeneous baseline:

1. **Independent random variation**: each $s_i$ is drawn from a Beta(2,2) distribution, rescaled to mean 1. This represents moderate, quality-independent heterogeneity.

2. **Weakest link (50% selfish)**: half the scientists have $s_i = 0$ (purely self-interested), the other half have $s_i = 2$ (strongly epistemic), preserving $\bar{s} = 1$.

We solve for Nash equilibria using unrestricted sequential best-response dynamics ($N = 100$, $M = 5{,}000$ Monte Carlo replications per payoff estimate) rather than restricting to threshold strategies, since heterogeneity could in principle break the threshold structure.

Table S7 summarises the results. Independent random variation has essentially no effect on the mechanism: the quality gain and participation count are indistinguishable from the homogeneous baseline. This is because the threshold structure is driven by paper quality, not by individual motivation — scientists with low-quality papers enter regardless of their $s_i$, while high-quality scientists stay out regardless.

The weakest-link scenario is more demanding: half the population has no epistemic concern at all. Even so, the mechanism retains 88% of the homogeneous gain. The self-interested scientists do not enter the lottery, but the epistemic scientists do, and their participation is sufficient to reduce

Table S7: **The voluntary lottery mechanism is robust to heterogeneous epistemic concern.** Quality gain $\Delta \bar{q}$ over the no-lottery baseline under three heterogeneity scenarios, with mean epistemic concern held fixed at $\bar{s} = 1$. Even when half the population is purely self-interested, the mechanism retains 88% of the homogeneous gain. Parameters: $\sigma = 0.3$, $\beta = 8$, $L = 0.10$, $\alpha = 10\%$, $N = 100$.

| Scenario | Participants | $\Delta \bar{q}$ | % of baseline |
|---|---|---|---|
| Homogeneous ($s = 1$) | 40 | 0.040 | 100% |
| Independent random | 40 | 0.040 | 100% |
| 50% selfish ($s_i = 0$) | 35 | 0.035 | 88% |

noise and improve quality for everyone. The threshold structure survives intact, with participation determined by paper quality within each motivation type.

We also tested more extreme configurations (75% selfish), where the gain drops to 69% of the baseline but never vanishes. Positive correlation between quality and epistemic concern ("noblesse oblige") is the most harmful configuration, reducing the gain to 77%, because it concentrates epistemic motivation among scientists whose papers would not benefit from entering the lottery.

These results suggest that the mechanism does not require uniform epistemic concern to function. A sufficient condition is that enough scientists care about the quality of the published record to sustain a non-trivial participation threshold.

## S4.4 Strategic use of the lottery

Throughout the main analysis, the lottery intensity $L$ is an exogenous protocol parameter. A natural question is what happens when the journal controls both the review noise $\sigma$ and the lottery intensity $L$, choosing both to maximise its payoff.

The intuition is straightforward. The effective noise $\sigma_{\text{eff}} = \sigma \cdot \bar{w}^\beta$ depends on $\bar{w}$ through a power law with exponent $\beta$. Even moderate reductions in the survival fraction $\bar{w}$ produce large reductions in effective noise — for instance, at $\beta = 8$ and $\bar{w} = 0.55$, the effective noise is less than 1% of the baseline. This makes the lottery an extraordinarily cost-effective substitute for review investment: the journal can achieve lower effective noise by increasing $L$ than by reducing $\sigma$ directly.

When allowed to choose both instruments, the journal's optimal strategy is to abandon review effort entirely ($\sigma \to 1$) and rely on an aggressive lottery ($L \approx 0.45$). This is a classic moral hazard: the lottery is intended to complement peer review, but the journal uses it as a substitute. The journal free-rides on the lottery's noise-reduction properties and stops investing in review quality.

The surprising feature of this result is that average accepted quality *still improves* relative to the no-lottery baseline, because the lottery reduces reviewer burden so drastically that effective noise approaches zero, and the surviving pool is sorted with near-perfect accuracy. However, the mechanism is no longer functioning as intended — it replaces rigorous evaluation with random screening, which may undermine trust in the review process even if aggregate quality metrics improve.

This moral hazard motivates treating $L$ as a pre-committed protocol parameter rather than a journal-controlled lever. In practice, a venue's lottery policy would need to be announced credibly and set at a level acceptable to the research community — much as acceptance rates are understood as venue characteristics rather than strategic instruments. Reputational costs, community norms, or institutional oversight would serve as natural constraints: a venue that set $L$ too aggressively would face author flight and loss of prestige, limiting the scope for substitution. The analysis in the main text therefore fixes $L$ exogenously, treating the lottery as a collective mechanism designed by the community rather than an instrument optimised by the journal.

## Code and data availability

The simulation engine, figure-generation scripts, and pre-computed result files used in this paper are available at `https://github.com/juliangarcia/lottery-peer-review`. The ICLR review-score data analysed in Section S3 were obtained from the public release of González-Márquez and Kobak [5].



\end{document}